\documentclass[conference]{IEEEtran}

\usepackage{cite}

\usepackage{amsmath}
\usepackage{mathrsfs}
\usepackage{bm}

\usepackage{algorithmic}

\usepackage{array}

\usepackage{caption}

\usepackage{url}
\usepackage{graphicx,amsmath,amssymb,amsfonts}
\usepackage{algorithmic,algorithm}

\usepackage{setspace}

\usepackage{xcolor}

\usepackage{epstopdf}

\newtheorem{theorem}{\textbf{Theorem}}

\newtheorem{lemma}{\textbf{Lemma}}

\newtheorem{corollary}{\textbf{Corollary}}

\renewcommand{\IEEEQED}{\IEEEQEDopen}

\makeatletter

\newcommand{\Rmnum}[1]{\expandafter\@slowromancap\romannumeral #1@}
\makeatother

\begin{document}
	\bstctlcite{ref:BSTcontrol}

    \title{QoS-Constrained Federated Learning Empowered by Intelligent Reflecting Surface}


	\author{\IEEEauthorblockN{
    		Jingheng~Zheng,
    		Wanli~Ni,
    		and Hui~Tian
    		\IEEEauthorblockA{
    			State Key Laboratory of Networking and Switching Technology \\
    			Beijing University of Posts and Telecommunications, Beijing 100876, China \\
    			Email: \{zhengjh, charleswall, tianhui\}@bupt.edu.cn
    		}
    	}
    \vspace{-0.8cm}
    }

\vspace{-0.8cm}
\maketitle	

\begin{abstract}
    This paper investigates the model aggregation process in an over-the-air federated learning (AirFL) system, where an intelligent reflecting surface (IRS) is deployed to assist the transmission from users to the base station (BS). With the purpose of overcoming the absence of the security examination against malicious individuals, successive interference cancellation (SIC) is adopted as a basis to support analyzing statistic characteristics of model parameters from devices. The objective of this paper is to minimize the mean-square-error by jointly optimizing the receive beamforming vector at the BS, transmit power allocation at users, and phase shift matrix of the IRS, subject to the transmit power constraint for devices, unit-modulus constraint for reflecting elements, SIC decoding order constraint and quality-of-service constraint. To address this complicated problem, alternating optimization is employed to decompose it into three subproblems, where the optimal receive beamforming vector is obtained by solving the first subproblem with the Lagrange dual method. Then, the convex relaxation method is applied to the transmit power allocation subproblem to find a suboptimal solution. Eventually, the phase shift matrix subproblem is addressed by invoking the semidefinite relaxation. Simulation results validate the availability of IRS and the effectiveness of the proposed scheme in improving federated learning performance.
\end{abstract}


\let\thefootnote\relax\footnotetext{This paper is funded by Beijing Univ. of Posts and Telecommun.-China Mobile Research Institute Joint Innovation Center.}

\section{Introduction}
The significant development of available computation resources on edge devices accelerates the legacy centralized machine learning (ML) framework evolving towards a distributed but collaborative manner~\cite{Mingzhe2021A, Jed2020User}. As one of the most promising edge learning algorithms, federated learning (FL) provides a distributed methodology to render the participants collaboratively train the shared global model while keeping their raw data locally.
Specifically, devices in FL system firstly take the advantage of their own original data to train the ML model locally, and upload the trained model parameters to the base station (BS) for global aggregation after local training~\cite{Wanli2020Federated}.
Nevertheless, the conventional transmit-then-compute approach of FL is becoming impractical owing to the inevitably high latency and low spectrum efficiency~\cite{Qiao2021Integrated}. Moreover, the undesirable wireless channel condition may deteriorate the performance of FL as well.

With the purpose of tackling the aforementioned problems, the over-the-air computation (AirComp) technique and intelligent reflecting surface (IRS) are regarded as promising remedies.
On the one hand, AirComp exploits the superposition property of the multiple access channels to compute nomographic functions~\cite{Guangxu2019MIMO}, which effectively reduce the aggregation latency at BS. Meanwhile, AirComp achieves improved spectrum efficiency through concurrent transmission from different devices over the same spectrum resource as well.
On the other hand, due to the ability of proactively adjusting the amplitudes and phase shifts of reflecting elements~\cite{Qingqing2019Intelligent}, IRS is competent to reconfigure the wireless environment and alleviate the deteriorative wireless channel conditions.

So far, both FL and IRS has been implemented in multifarious existing works like~\cite{Mingzhe2021A} and~\cite{Wanli2020Federated}. In~\cite{Mingzhe2021A}, the authors built an explicit relationship between packet error rate and FL performance and minimized the FL loss function by optimizing resource allocation and user selection. By deploying multiple IRSs in FL system, Ni \emph{et. al}~\cite{Wanli2020Federated} developed a novel framework of resource allocation and device selection for FL system.
For AirComp, there has been a quite number of research works such as~\cite{Qiao2021Integrated},~\cite{Guangxu2019MIMO} and~\cite{Li2020Computation}. Concretely speaking, in order to integrate the sensing, computation and communication in beyond fifth-generation internet of things networks, Qi \emph{et. al}~\cite{Qiao2021Integrated} respectively minimized the mean-square-error (MSE) and maximized the weighted sum-rate through the joint designing of transmit and receive beamforming matrices. With the aim of minimizing MSE of the received signal and designing the channel feedback, Zhu \emph{et. al}~\cite{Guangxu2019MIMO} invoked a differential geometry based approach to optimize the receive beamforming matrix and established a novel AirComp-multicasting duality. Considering an analogous concept of AirComp, i.e., computation over multiple access channel, Chen \emph{et. al}~\cite{Li2020Computation} optimized the transceiver to maximize the derived achievable function rate and a low-complexity signaling procedure is proposed as well.
However, the above works overemphasize the efficiency of the aggregation process in FL system but pay less attention to decode individual model parameters, which acts as a basis for detecting malicious FL participants. It is of significance to decode and analyze statistic characteristics of the uploaded model parameters to distinguish malicious devices.

In this paper, we focus on improving the performance of AirComp based FL as known as over-the-air FL (AirFL)~\cite{Wanli2021integrating} while ensuring the quality-of-service (QoS) constraint. Since AirComp without individual information decoding can be regarded as a particular case of non-orthogonal multiple access (NOMA) techniques~\cite{Dongzhu2020Privacy}, then the successive interference cancellation (SIC) can be employed as the decoding methodology. The main contribution of this paper can be summarized as follows:
1) A non-convex MSE minimization problem with QoS constraint is formulated for the AirFL system assisted by IRS, where the SIC technique is adopted as the decoding methodology.
2) To solve the intractable problem, an alternating optimization based algorithm is proposed to design the receive beamforming vector, transmit power allocation and phase shift matrix in an iterative manner.
3) Simulation results validate the availability of IRS and the effectiveness of the proposed scheme in improving AirFL performance.
%
%


\section{System Model And Problem Formulation} \label{system model and problem formulation}

\subsection{System Model} \label{system model}
\begin{figure} [t]
	\centering
	\includegraphics[scale=0.7]{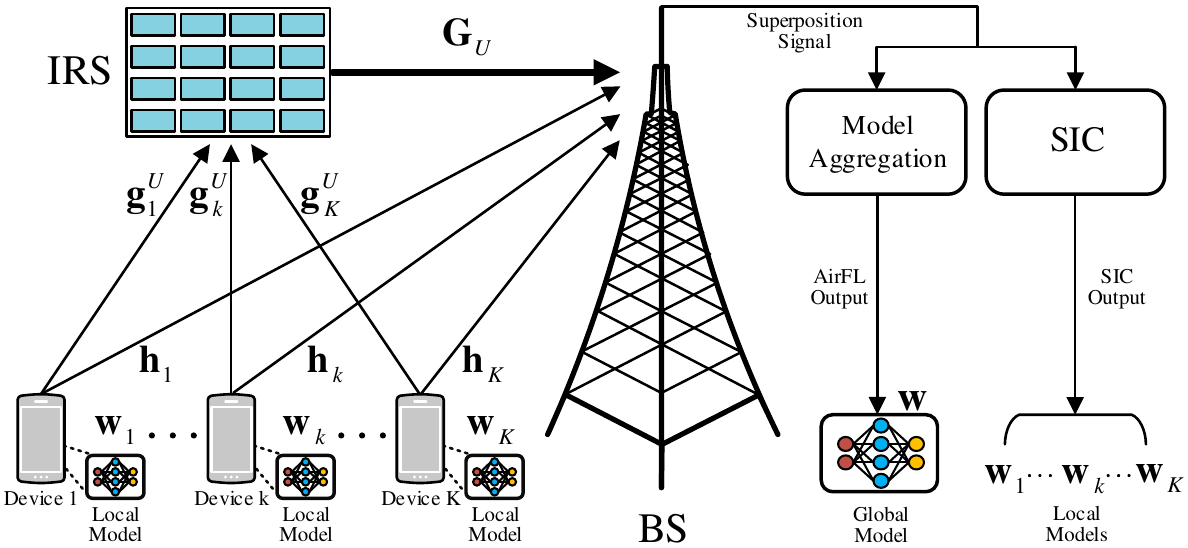}
	\caption{System model of IRS-empowered AirFL.}
	\label{system_model}
\end{figure}

As depicted in Fig.~\ref{system_model}, an IRS-empower AirFL system is considered, which is comprised of one BS, $K$ devices as well as one IRS. Devices are indexed by ${\mathcal{K}} = \left\{ {1,...,K} \right\}$ and $M$ reflecting elements of IRS are indexed by $\mathcal{M} = \left\{ {1,...,M} \right\}$. Let the $M \times M$-dimensional diagonal matrix ${\bf{\Theta }} = {\rm diag}\left\{ {{e^{j{\phi _1}}},...,{e^{j{\phi _M}}}} \right\}$ denotes the phase shift matrix of IRS, where ${\phi _m} \in \left[ {0,2\pi } \right]$ refers to the phase shift provided by the $m$-th reflecting element. ${{\bf{W}}_k}$ denotes the local model parameters of the $k$-device learnt from its local dataset. With the purpose of improving spectrum efficiency and reduce latency, AirComp is adopted to enable the concurrent transmission of all devices over the same bandwidth.

In the scenario of AirFL, the BS performs different aggregation process for all ${{\bf{W}}_k}$ uploaded by devices through taking the advantage of variant nomographic functions~\cite{Li2020Computation} presented as follows
\begin{equation}
    {\bf{W}} = \psi \left( {\sum\limits_{k = 1}^K {{\varphi _k}\left( {{{\bf{W}}_k}} \right)} } \right),
\end{equation}
where $\bf{W}$ denotes the aggregated global model parameters, ${\varphi _k}\left(  \cdot  \right)$ and $\psi \left(  \cdot  \right)$ respectively represent the pre-process function of the $k$-th device and the post-process function at BS.

After pre-processing, ${{{\bf{W}}_k}}$ is transformed into transmitting symbol ${s_k} \in \mathbb{C}$ and it is assumed to be an independently and identically distributed random variable with zero mean and unit variance. It is further assumed that each device equips with single antenna and BS is equipped with ${N_r}$ antennas to receive signals from devices. In addition, let $s = \sum\nolimits_{k = 1}^K {{s_k}}$ denote the desired superposition signal. Based on these assumptions, the estimated superposition signal at BS, i.e., ${\hat s}$, can be expressed as follows

\begin{equation}
    \hat s = {{\bf{b}}^H}\left[ {\sum\limits_{k = 1}^K {\left( {{{\bf{h}}_k} + {\bf{G}}_U^H{\bf{\Theta g}}_k^U} \right)\sqrt {{p_k}} {s_k} + {\bf{n}}} } \right],
\end{equation}
where ${\bf{x}}^H$ stands for the conjugate transpose of $\bf{x}$, ${\bf{b}} \in {\mathbb{C}^{{N_r} \times 1}}$ denotes the receive beamforming vector of BS, ${{p_k}}$ refers to transmit power of the $k$-th device, ${\bf{h}}_k \in \mathbb{C}^{{N_r} \times 1}$ is the Rician channel vector between the $k$-th device and BS, ${\bf{g}}_k^U \in {\mathbb{C}^{M \times 1}}$ represents the Rayleigh channel from the $k$-th device to IRS and
${{{\bf{G}}}_U} \in {\mathbb{C}^{M \times {N_r}}}$ denotes the Rician channel matrix between IRS and BS. ${\bf{n}} \in \mathbb{C}^{{N_r} \times 1}$ is the additive white Gaussian noise (AWGN) vector with each element independently distributed as $CN\left( {0,{\sigma ^2}} \right)$, where $\sigma ^2$ is the noise power. For the sake of notational convenience, let ${{{\bf{\bar h}}}_k} = {{\bf{h}}_k} + {{\bf{G}}_U^H{\bf{\Theta g}}_k^U}$.

We take the advantage of MSE to quantify the performance of AirFL in the sense that alleviating signal distortion results in the improvement of federated learning. Concretely speaking, distortion of ${\hat s}$ with respect to $s$ is measured by MSE defined as follows
\begin{equation}
{\rm MSE}\left( {\hat s,s} \right)
\!=\! \mathbb{E} ( {{{\left| {\hat s - s} \right|}^2}} )
\!=\! \sum\limits_{k = 1}^K {{{\left| {{{\bf{b}}^H}{{{\bf{\bar h}}}_k}\sqrt {{p_k}} \!-\! 1} \right|}^2}}  \!+\! {{{\left\| {\bf{b}} \right\|}^2}}{\sigma ^2},
\end{equation}
where $\left\| {\bf{x}} \right\|$ represents the 2-norm of vector $\bf{x}$ and $\left| x \right|$ denotes the modulus of complex number $x$.

Although AirFL technique can straightforwardly aggregate local model parameters from participant devices, decoding and recording them are still imperative for security assurance techniques against malicious devices. Therefore, as depicted in Fig. 1, SIC is adopted to decode model parameters of each device from the received superposition signal. In this way, BS is capable of distinguishing malicious devices who mount attacks on the system via analyzing statistic characteristics of the decoded model parameters~\cite{Nguyen2021Privacy}. Beyond that, memorizing aggregated global model parameters benefits the convergence of the federated learning process through techniques like momentum based stochastic gradient descent~\cite{Yujun2020Deep} as well. Specifically, the uplink data rate of the $k$-th device, ${R_k}$, under SIC decoding scheme is denoted by
\begin{equation}
    {R_k} = B{\log _2}\left( {1 + \frac{{{{\left| {{{\bf{b}}^H}{{{\bf{\bar h}}}_k}} \right|}^2}{p_k}}}{{\sum\nolimits_{\pi \left( {k'} \right) > \pi \left( k \right)} {{{\left| {{{\bf{b}}^H}{{{\bf{\bar h}}}_{k'}}} \right|}^2}{p_{k'}}}  + {{\left\| {\bf{b}} \right\|}^2}{\sigma ^2}}}} \right),
\end{equation}
where ${\pi \left( k \right)}$ refers to the decoding order of the $k$-th device (e.g., $\pi \left( k \right) = 3$ means that $k$-th device is the third one to be decoded at BS) and $B$ denotes the available bandwidth.

Strictly speaking, in the uplink transmission scenario, the $k$-th device is decoded ahead of the ${k'}$-th device when ${\left| {{{\bf{b}}^H}{{{\bf{\bar h}}}_k}} \right|^2}{p_k} > {\left| {{{\bf{b}}^H}{{{\bf{\bar h}}}_{k'}}} \right|^2}{p_{_{k'}}}$ is satisfied. However, since jointly optimizing ${\bf{b}}$ and ${p_k}$ to determine the decoding order remains extremely complicated, we alternatively utilize ${\left\| {{{{\bf{\bar h}}}_k}} \right\|^2}$ to derive ${\pi \left( k \right)}$ for simplicity. Hence, the $k$-th device is previously decoded than the ${k'}$-th device if ${\left\| {{{{\bf{\bar h}}}_k}} \right\|^2} > {\left\| {{{{\bf{\bar h}}}_{k'}}} \right\|^2}$ and the following group of constraints should be satisfied as well to successfully perform SIC~\cite{MD2016Dynamic}, i.e.,
\vspace{-0.4cm}
\begin{equation}
{\left| {{{\bf{b}}^H}{{{\bf{\bar h}}}_k}} \right|^2}{p_k} - \sum\limits_{k' = k + 1}^K {{{\left| {{{\bf{b}}^H}{{{\bf{\bar h}}}_{k'}}} \right|}^2}{p_{k'}}}  \ge {p_{\rm gap}},\forall k \in \mathcal{K}\backslash \{ K\},
\end{equation}
where $p_{\rm gap}$ refers to the minimum processed transmit power difference between the signal to be decoded and those remain non-decoded. To facilitate expression, we rearrange the indexes of devices such that the device with smaller index are decoded first henceforth.

\subsection{Problem Formulation} \label{problem_formulation}
In this paper, we aim at optimizing the performance of the AirFL system empowered by IRS while employing SIC technique to decode the uploaded model parameters from the superposition signal. Due to the fact that achieving joint optimization of MSE and QoS is intractable, we alternatively make the objective of minimizing MSE while ensuring the QoS constraint, i.e., the data rate constraint. As a result, the optimization problem can be formulated as follows
\begin{subequations}
    \begin{eqnarray}
        \mathcal{P}1:
        &\mathop {\min }\limits_{{\bf{p}},{\bf{\Theta }},{\bf{b}}} & {\rm MSE}\left( {\hat s,s} \right) \\
        &s.t.&0 < {p_k} \le {P_{\max }}, \forall k \in {\mathcal{K}}, \\
        &{}&0 \le {\phi _m} \le 2\pi , \forall m \in \mathcal{M}, \\
        &{}&{R_k} \ge {R_{\min }}, \forall k \in {\mathcal{K}}, \\
        &{}&(5),
    \end{eqnarray}
\end{subequations}
where ${\bf{p}} = {\left[ {{p_1},{p_2},...,{p_K}} \right]^T}$ denotes the transmit power vector, ${P_{\max }}$ is the maximum transmit power of each device and ${R_{\min}}$ represents the minimum data rate requirement.

The coupling of optimization variables in the objective and constraints renders $\mathcal{P}1$ a non-linear and non-convex problem. It is still lack of standard methodologies to straightforwardly cope with these problems. With the purpose of making it tractable, we then adopt alternating optimization (AO) to decompose problem $\mathcal{P}1$ into a sequence of subproblems and find an suboptimal solution in an iterative manner.

\section{Proposed Algorithm} \label{proposed algorithm}
\subsection{Receive Beamforming Vector} \label{Receive Beamforming Vector}
When both transmit power allocation $\bf{p}$ and phase shift matrix $\bf{\Theta }$ are fixed, preliminary problem $\mathcal{P}1$ can be transformed into the subproblem of optimizing receive beamforming vector $\bf{b}$ subjects to constraints (6d) and (5), which is given by
\begin{subequations}
    \begin{eqnarray}
        \mathcal{P}2:
        &\mathop {\min }\limits_{{\bf{b}}} &\sum\limits_{k = 1}^K {{{\left| {{{\bf{b}}^H}{{{\bf{\bar h}}}_k}\sqrt {{p_k}}  - 1} \right|}^2}}  + {{{\left\| {\bf{b}} \right\|}^2}}{\sigma ^2} \\
        &s.t.&{\left| {{{\bf{b}}^H}{{{\bf{\bar h}}}_k}} \right|^2}{p_k} - {\gamma _{\min }}\sum\limits_{k' = k + 1}^K {{{\left| {{{\bf{b}}^H}{{{\bf{\bar h}}}_{k'}}} \right|}^2}{p_{k'}}} \nonumber \\
 &&- {\gamma _{\min }}{\left\| {\bf{b}} \right\|^2}{\sigma ^2} \ge 0, \forall k \in {\mathcal{K}}, \\
        &{}& (5),
    \end{eqnarray}
\end{subequations}
where ${\gamma _{\min }} = {2^{{{{R_{\min }}} \mathord{\left/ {\vphantom {{{R_{\min }}} B}} \right. \kern-\nulldelimiterspace} B}}} - 1$ denotes the minimum signal-to-interference-plus-noise ratio to achieve $R_{\min }$. Since $\mathcal{P}2$ is non-convex with respect to (w.r.t.) $\bf{b}$ and is hard to be addressed directly, we intent to obtain the solution through invoking the Lagrange dual method.

For the sake of notational convenience, we respectively define auxiliary matrices ${\bf{H}}_k$, ${\bf{A}}_k$ and ${\bf{B}}_k$ as follows
\begin{eqnarray}
&{{\bf{H}}_k} = {{{\bf{\bar h}}}_k}{\bf{\bar h}}_k^H, \forall k \in {\mathcal{K}}, \\
&{{\bf{A}}_k} = {p_k}{{\bf{H}}_k} - {\gamma _{\min }}\sum\limits_{k' = k + 1}^K {{p_{k'}}{{\bf{H}}_{k'}}}  - {\gamma _{\min }}{\sigma ^2}{\bf{I}}, \forall k \in {\mathcal{K}}, \\
&{{\bf{B}}_k} = {p_k}{{\bf{H}}_k} - \sum\limits_{k' = k + 1}^K {{p_{k'}}{{\bf{H}}_{k'}}}, \forall k \in \mathcal{K}\backslash \{ K\}.
\end{eqnarray}
As a result, problem $\mathcal{P}2$ can be equivalently rewritten as
\begin{subequations}
    \begin{eqnarray}
        \mathcal{P}2.1:
        &\mathop {\min }\limits_{{\bf{b}}} &\sum\limits_{k = 1}^K {{{\left| {{{\bf{b}}^H}{{{\bf{\bar h}}}_k}\sqrt {{p_k}}  - 1} \right|}^2}}  + {\left\| {\bf{b}} \right\|^2}{\sigma ^2} \\
        &s.t.&{{\bf{b}}^H}{{\bf{A}}_k}{\bf{b}} \ge 0, \forall k \in \mathcal{K}, \\
        &{}&{{\bf{b}}^H}{{\bf{B}}_k}{\bf{b}} - {p_{\rm gap}} \ge 0, \forall k \in \mathcal{K}\backslash \{ K\}.
    \end{eqnarray}
\end{subequations}
Furthermore, the Lagrangian function of problem $\mathcal{P}2.1$ can be obtained as follows
\begin{align}
&{\cal L}\left( {{\bf{b}},\left\{ {{\lambda _k}} \right\},\left\{ {{\mu_k}} \right\}} \right) = K + {p_{\rm gap}}\sum\limits_{k = 1}^K {{\mu _k}} - {{\bf{b}}^H}\left( {\sum\limits_{k = 1}^K {{{{\bf{\bar h}}}_k}\sqrt {{p_k}} } } \right) \notag  \\
&+ {{\bf{b}}^H}\left( {\sum\limits_{k = 1}^K {{\bf{\bar H}}_k + {\sigma ^2}{\bf{I}}} } \right){\bf{b}} - \left( {\sum\limits_{k = 1}^K {{\bf{\bar h}}_k^H\sqrt {{p_k}} } } \right){\bf{b}},   
\end{align}
where $\{\lambda_k\}$ and $\{\mu_k\}$ are non-negative Lagrange multipliers, ${\bf{\bar H}}_k ={p_k}{{\bf{H}}_k} - {\lambda _k}{{\bf{A}}_k} - {\mu _k}{{\bf{B}}_k}$ and $\bf{I}$ denotes the identity matrix. Note that the lack of the $K$-th constraint in (11c) makes the related expressions less concise. In order to deal with this issue, we artificially make ${\mu _K}=0$ and ${{\bf{B}}_K} = {p_K}{{\bf{H}}_K}$ to facilitate the expression of summation terms related to $\left \{ {\mu_k} \right \}$.

By means of taking the first order partial derivative of $\mathcal{L}\left( {{\bf{b}},\left\{ {{\lambda _k}} \right\},\left\{ {{\mu _k}} \right\}} \right)$ w.r.t. $\bf{b}$ and forcing the result equal to zero, the optimal receive beamforming vector ${\bf{b}}^*$ is obtained as

\begin{equation}
{{\bf{b}}^*} = \left( \sum\limits_{k = 1}^K  {\bf{\bar H}}_k + {\sigma ^2} \bf{I} \right)^{ - 1} \left( \sum\limits_{k = 1}^K {\bf{\bar h}}_k \sqrt{p_k} \right),
\end{equation}
where ${\bf{X}^{-1}}$ denotes the inverse matrix of matrix ${\bf{X}}$.
The design of the receive beamforming vector $\bf{b}$ is also known as the minimum mean-square-error (MMSE) criterion.

\begin{figure*}
	\begin{subequations}
		\begin{eqnarray}
		\mathcal{P}2.2:
		&\mathop {\max }\limits_{{\left\{ {{\lambda _k}} \right\}},{\left\{ {{\mu _k}} \right\}}} &- \left( {\sum\limits_{k = 1}^K {{\bf{\bar h}}_k^H\sqrt {{p_k}} } } \right){\left\{ {\sum\limits_{k = 1}^K {{p_k}{{\bf{H}}_k}} \left[ {1 - {\lambda _k} - {\mu _k} + \sum\limits_{i = 1}^{k - 1} {\left( {{\lambda _i}{\gamma _{\min }} + {\mu _i}} \right)} } \right] + \left[ {\left( {\sum\limits_{k = 1}^K {{\lambda _k}} } \right){\gamma _{\min }} + 1} \right]{\sigma ^2}{\bf{I}}} \right\}^{ - 1}} \nonumber \\
		&&*\left( {\sum\limits_{k = 1}^K {{{{\bf{\bar h}}}_k}\sqrt {{p_k}} } } \right) + K + {p_{\rm gap}}\sum\limits_{k = 1}^K {{\mu _k}} \\
		&s.t.&{\lambda _k} \ge 0,{\mu _k} \ge 0, \forall k \in \mathcal{K}.
		\end{eqnarray}
		\hrulefill
	\end{subequations}
\end{figure*}

The Lagrange dual problem of $\mathcal{P}2.1$ can be reformulated as $\mathcal{P}2.2$ presented at the top of the next page via substituting ${\bf{b}}^*$ into (12). Since the inverse matrix is fairly complicated, it is intractable to derive closed-form expressions for Lagrangian multipliers $\left\{ {{\lambda _k}} \right\}$ and $\left\{ {{\mu _k}} \right\}$. Thus, we alternatively adopt sub-gradient method to update $\left\{ {{\lambda _k}} \right\}$ and $\left\{ {{\mu _k}} \right\}$ in an iterative manner.
It has been proved in~\cite{Boyd2004Convex} that sub-gradient method with constant step size is convergent, so we respectively update $\left\{ {{\lambda _k}} \right\}$ and $\left\{ {{\mu _k}} \right\}$ in accordance with the following rules:
\begin{align}
\lambda _k^{\left( {t + 1} \right)} = {\left[ {\lambda _k^{\left( t \right)} - {\delta _1}\left( {{{\bf{b}}^{\left( t \right)}}^H{{\bf{A}}_k}{{\bf{b}}^{\left( t \right)}}} \right)} \right]^ + }, \forall k \in \mathcal{K},
\end{align}
\begin{align}
\mu _k^{\left( {t + 1} \right)} = {\left[ {\mu _k^{\left( t \right)} - {\delta _2}\left( {{{\bf{b}}^{\left( t \right)}}^H{{\bf{B}}_k}{{\bf{b}}^{\left( t \right)}} - {p_{\rm gap}}} \right)} \right]^ + }, \forall k \in \mathcal{K}\backslash \{ K\},
\end{align}
where $\delta _1$ and $\delta _2$ are constant step length, ${\left[ x \right]^ + } = \max \left\{ {0,x} \right\}$ and ${{\bf{b}}^{\left( t \right)}}$ denotes the obtained value of $\bf{b}$ at the $t$-th iteration. The Lagrange dual method based algorithm for obtaining receive beamforming vector is summarized in {\bf{Algorithm 1}}, where $T_{1,\max}$ refers to the maximum iteration number and $\varepsilon_1$ is an adjustable convergence accuracy.

\subsection{Transmit Power Allocation} \label{Transmit Power Allocation}

After the receive beamforming vector $\bf{b}$ is obtained, we attempt to solve transmit power allocation subproblem with fixed phase shift matrix $\bf{\Theta }$. In accordance with $\mathcal{P}1$, we ignore irrelevant terms and the transmit power allocation subproblem can be reduced to
\begin{subequations}
    \begin{eqnarray}
        \mathcal{P}3:
        &\mathop {\min }\limits_{{\bf{p}}} &\sum\limits_{k = 1}^K {{{\left| {{{\bf{b}}^H}{{{\bf{\bar h}}}_k}} \right|}^2}{p_k} - 2{\mathop{\rm Re}\nolimits} \left\{ {{{\bf{b}}^H}{{{\bf{\bar h}}}_k}} \right\}\sqrt {{p_k}} }    \\
        &s.t.&(5),~(6b)~{\rm and}~(7b),
    \end{eqnarray}
\end{subequations}

Though all constraints of $\mathcal{P}3$ are linear functions of ${p_k}$, the objective is still non-convex w.r.t. them. With the purpose of addressing its non-convexity, we introduce an auxiliary variable ${\eta _k}$ that satisfies constraint ${\eta _k} = \sqrt {{p_k}}$ to transform $\mathcal{P}3$ into a convex optimization problem. Since equality constraint usually makes it intractable to solve, we simply relax it to an inequality constraint and the relaxed problem is given by
\begin{subequations}
    \begin{eqnarray}
        \mathcal{P}3.1:
        &\mathop {\min }\limits_{{\bf{p}},\{{\eta _k}\}} &\sum\limits_{k = 1}^K {{{\left| {{{\bf{b}}^H}{{{\bf{\bar h}}}_k}} \right|}^2}\eta _k^2 - 2{\mathop{\rm Re}\nolimits} \left\{ {{{\bf{b}}^H}{{{\bf{\bar h}}}_k}} \right\}{\eta _k}}~  \\
        &s.t.&{\eta _k} \le \sqrt {{p_k}}, \forall k \in \mathcal{K},\\
        &{}&(5),~(6b)~{\rm and}~(7b).
    \end{eqnarray}
\end{subequations}

Obviously, $\mathcal{P}3.1$ is a joint convex problem of ${\bf{p}}$ and $\{{\eta _k}\}$, which can be solved by existing convex toolkits such as CVX. Note that the allowed maximum iteration number is limited to ${T_{2,\max }}$ and the obtained ${\bf{p}^*}$ might be suboptimal due to the relaxation.

\begin{algorithm}[t]
	\caption{Lagrange Dual Method Based Algorithm for Receive Beamforming Vector.}
	\label{Lagrange Dual Method Based Algorithm for Receive Beamforming Vector}
	\begin{algorithmic}[1]
	\STATE {\bf Input:}
	${{\gamma _{\min }}}$, ${\sigma ^2}$, $p_{\rm gap}$, $\delta _1$, $\delta _2$, $T_{1,max}$, $\varepsilon_1$, $\{{{\bf{\bar h}}}_k\}$, $\{p_k\}$, $\{{\lambda _k}\}$, $\{{\mu _k}\}$, $t$.\\
	\STATE {\bf Output:}
	the receive beamforming vector ${\bf{b}}^*$.	
	\STATE Initialize ${{\bf{b}}^{\left( 0 \right)}}$,$\left\{ {\lambda _k^{\left( 0 \right)}} \right\}$, $\left\{ {\mu _k^{\left( 0 \right)}} \right\}$, $\delta _1$, $\delta _2$, $T_{1,max}$, $\varepsilon_1$, $t=0$.
	\WHILE {$t < {T_{1,max }}$ \&\& $\left\| {{{\bf{b}}^{\left( t \right)}} - {{\bf{b}}^{\left( {t - 1} \right)}}} \right\| > \varepsilon_1 $ }
	\STATE $t=t+1$.
	\STATE Use $\left\{{\lambda _k^{\left( t-1 \right)}}\right\}$ and $\left\{{\mu _k^{\left( t-1 \right)}}\right\}$ to calculate ${{{\bf{b}}^{\left( t \right)}}}$ via (13).
	\STATE Use ${{{\bf{b}}^{\left( t \right)}}}$ to update $\left\{{\lambda _k^{\left( t \right)}}\right\}$ and $\left\{{\mu _k^{\left( t \right)}}\right\}$ via (15) and (16), respectively.
	\ENDWHILE
	\STATE ${{\bf{b}}^*} = {{\bf{b}}^{\left( t \right)}}$.
	\STATE {\bf Return:} the receive beamforming vector ${\bf{b}}^*$.
	\end{algorithmic}
\end{algorithm}

\subsection{Phase Shift Matrix} \label{Phase Shift Matrix}

When receive beamforming vector $\bf{b}$ and power allocation $\bf{p}$ are obtained, we aim at deriving the phase shift matrix ${\bf{\Theta }}$ for IRS. The subproblem of optimizing ${\bf{\Theta }}$ is given by
\begin{subequations}
    \begin{eqnarray}
        \mathcal{P}4:
        &\mathop {\min }\limits_{{\bf{\Theta }}} &\sum\nolimits_{k = 1}^K {{{\left| {{{\bf{b}}^H}{{{\bf{\bar h}}}_k}\sqrt {{p_k}}  - 1} \right|}^2}}  + {\left\| {\bf{b}} \right\|^2}{\sigma ^2}  \\
        &s.t.&(5),~(6c)~{\rm and}~(6d),
    \end{eqnarray}
\end{subequations}
where ${\bf{\Theta }}$ is implicitly contained in ${{{\bf{\bar h}}}_k}$.

With the purpose of explicitly expressing the optimization variables and making the notation more concise, we first let ${\bf{v}} = {\left[ {{e^{j{\phi _1}}},{e^{j{\phi _2}}},...,{e^{j{\phi _M}}}} \right]^T}$ and ${{\bf{D}}_k} = {\bf{G}}_U^Hdiag\left\{ {{\bf{g}}_k^U} \right\}$, which make ${{{\bf{\bar h}}}_k} = {{\bf{h}}_k} + {{\bf{D}}_k}{\bf{v}}$ in consequence. Since the expression of subproblem remains complicated, we further make ${{\bf{\Phi }}_k} = {\bf{D}}_k^H{{\bf{b}}}\sqrt {{p_k}}$ and ${\rho _k} = {\bf{h}}_k^H{\bf{b}}\sqrt {{p_k}}$ to define the following expressions to simplify the subproblem
\begin{align}
{{\bm{\alpha }}_k} &= \sum\nolimits_{k' = 1}^K {{{\bf{\Phi }}_{k'}}\rho _{k'}^H - {{\bf{\Phi }}_k}}, \forall k \in \mathcal{K},\\
{{\bm{\beta }}_k} &= {{\bf{\Phi }}_k}\rho _k^H - {\gamma _{\min }}\sum\nolimits_{k' = k + 1}^K {{{\bf{\Phi }}_{k'}}\rho _{k'}^H}, \forall k \in \mathcal{K}, \\
{{\bm{\omega }}_k} &= {{\bf{\Phi }}_k}\rho _k^H - \sum\nolimits_{k' = k + 1}^K {{{\bf{\Phi }}_{k'}}\rho _{k'}^H}, \forall k \in \mathcal{K}\backslash \{ K\}.
\end{align}

Based on the above definitions, $\mathcal{P}4$ can be transformed to $\mathcal{P}4.1$ located at the top of next page, where ${C_{1,k}} = {\gamma _{\min }}\sum\nolimits_{k' = k + 1}^K {{{\left\| {{\rho _{k'}}} \right\|}^2}}  + {\gamma _{\min }}{\sigma ^2}{\left\| {\bf{b}} \right\|^2} - {\left\| {{\rho _k}} \right\|^2}$ and ${C_{2,k}} = \sum\nolimits_{k' = k + 1}^K {{{\left\| {{\rho _{k'}}} \right\|}^2}}  + {p_{\rm gap}} - {\left\| {{\rho _k}} \right\|^2}$. Intuitively, $\mathcal{P}4.1$ is a inhomogeneous quadratically constrained quadratic program problem but it is still non-convex w.r.t. ${\bf{v}}$.

\begin{figure*}
\begin{subequations}
	\begin{eqnarray}
	\mathcal{P}4.1:
	&\mathop {\min }\limits_{{\bf{v}}} &\sum\limits_{k = 1}^K {{{\bf{v}}^H}\left( {{{\bf{\Phi }}_k}{\bf{\Phi }}_k^H} \right){\bf{v}} + {{\bf{v}}^H}{{\bm{\alpha }}_k} + {\bm{\alpha }}_k^H{\bf{v}}}  \\
	&s.t.&\left| {{v_m}} \right| = 1, \forall m \in \mathcal{M},\\
	&{}&{{\bf{v}}^H}\left( {{{\bf{\Phi }}_k}{\bf{\Phi }}_k^H - {\gamma _{\min }}\sum\limits_{k' = k + 1}^K {{{\bf{\Phi }}_{k'}}{\bf{\Phi }}_{k'}^H} } \right){\bf{v}} + {{\bf{v}}^H}{{\bm{\beta }}_k} + {\bm{\beta }}_k^H{\bf{v}} \ge {C_{1,k}}, \forall k \in \mathcal{K}, \\
	&{}&{{\bf{v}}^H}\left( {{{\bf{\Phi }}_k}{\bf{\Phi }}_k^H - \sum\limits_{k' = k + 1}^K {{{\bf{\Phi }}_{k'}}{\bf{\Phi }}_{k'}^H} } \right){\bf{v}} + {{\bf{v}}^H}{{\bm{\omega }}_k} + {\bm{\omega }}_k^H{\bf{v}} \ge {C_{2,k}}, \forall k \in \mathcal{K}\backslash \{ K\}.
	\end{eqnarray}
	\vspace{0.2em}
	\hrulefill
\end{subequations}
\end{figure*}

To handle this problem, we homogenize it and take the advantage of matrix lifting technique to transform it into an semidefinite programming (SDP) problem. Specifically, we introduce an extra variable $t$ which satisfies $t^2=1$ to expand $\bf{v}$ to ${\bf{\bar v}} = {\left[ {{e^{j{\phi _1}}},{e^{j{\phi _2}}},...,{e^{j{\phi _M}}},t} \right]^T}$ and make ${\bf{V}} = {\bf{\bar v}}{{{\bf{\bar v}}}^H}$. Furthermore, we define several auxiliary matrices as
\begin{eqnarray}
{{\bf{F}}_0} = \sum\nolimits_{k = 1}^K {\left( {\begin{array}{*{20}{c}} {{{\bf{\Phi }}_k}{\bf{\Phi }}_k^H}&{{{\bm{\alpha }}_k}}\\ {{\bm{\alpha }}_k^H}&0 \end{array}} \right)},
\end{eqnarray}
\vspace{-6 mm}
\begin{align}
{{\bf{F}}_{1,k}} &= \left( {\begin{array}{*{20}{c}} {{{\bf{\Phi }}_k}{\bf{\Phi }}_k^H - {\gamma _{\min }}\sum\limits_{k' = k + 1}^K {{{\bf{\Phi }}_{k'}}{\bf{\Phi }}_{k'}^H} }&{{{\bm{\beta }}_k}}\\{{\bm{\beta }}_k^H}&0 \end{array}} \right), \forall k \in \mathcal{K}, \\
{{\bf{F}}_{2,k}} &= \left( {\begin{array}{*{20}{c}} {{{\bf{\Phi }}_k}{\bf{\Phi }}_k^H - \sum\limits_{k' = k + 1}^K {{{\bf{\Phi }}_{k'}}{\bf{\Phi }}_{k'}^H} }&{{{\bm{\omega }}_k}}\\{{\bm{\omega }}_k^H}&0 \end{array}} \right), \forall k \in \mathcal{K}\backslash \{ K\}.
\end{align}

Based on the defined matrices, $\mathcal{P}4.1$ can be equivalently rewritten as the following homogeneous SDP problem
\begin{subequations}
    \begin{eqnarray}
        \mathcal{P}4.2:
        &\mathop {\min }\limits_{{\bf{V}}} &{\rm Tr}\left( {{{\bf{F}}_0}{\bf{V}}} \right)  \\
        &s.t.&{\left[ {\bf{V}} \right]_{m,m}} = 1, \forall m \in \mathcal{M} \cup \{ M+1\}, \\
        &{}&Tr\left( {{{\bf{F}}_{1,k}}{\bf{V}}} \right) \ge {C_{1,k}}, \forall k \in \mathcal{K}, \\
        &{}&Tr\left( {{{\bf{F}}_{2,k}}{\bf{V}}} \right) \ge {C_{2,k}}, \forall k \in \mathcal{K}\backslash \{ K\}, \\
        &{}&{\bf{V}} \succeq 0, \\
        &{}&{\rm rank}({\bf{V}}) = 1,
    \end{eqnarray}
\end{subequations}
where ${\rm Tr}\left( {\bf{X}} \right)$ denotes the trace of matrix ${\bf{X}}$. Nevertheless, $\mathcal{P}4.2$ is still non-convex due to the non-convex constraint (27f). We employ semidefinite relaxation technique~\cite{Zhi2010Semidefinite} to cope with the non-convexity of $\mathcal{P}4.2$, which simply drops (27f) from $\mathcal{P}4.2$ to make it a convex problem as follow
\begin{subequations}
    \begin{eqnarray}
        \mathcal{P}4.3:
        &\mathop {\min }\limits_{{\bf{V}}} &{\rm Tr}\left( {{{\bf{F}}_0}{\bf{V}}} \right)  \\
        &s.t.&(27b),~(27c),~(27d),~(27e).
    \end{eqnarray}
\end{subequations}

Since $\mathcal{P}4.3$ is convex, it can be solved through CVX toolkit as well and the solution accuracy is set to be ${{\varepsilon _3}}$. Note that the obtained ${\bf{V}}$ may violate constraint (27f), so it is necessary to perform approximation techniques such as eigenvalue decomposition for ${\bf{V}}$ to obtain an approximate but feasible solution. Concretely speaking, we make ${\bf{\bar v}} \approx \sqrt {{\lambda _{\max }}} {\bf{q}}$ if ${\rm rank}({\bf{V}}) > 1$, where ${{\lambda _{\max }}}$ denotes the maximum eigenvalue of ${\bf{V}}$ and ${\bf{q}}$ represents its corresponding eigenvector. Finally, the optimized ${\bf{\Theta }^*}$ is recovered from ${\bf{\bar v}}$ by dropping the last element and diagonalizing the remaining elements.

\subsection{Alternating Optimization Based Algorithm} \label{Alternating Optimization Based Algorithm}

The overall AO based algorithm for solving problem $\mathcal{P}1$ is summarized in {\bf{Algorithm 2}}, where ${{\bf{b}}^{\left( t \right)}}$, ${{\bf{p}}^{\left( t \right)}}$, ${{\bf{\Theta }}^{\left( t \right)}}$ and ${\rm MSE}^{\left( t \right)}$ respectively represent the obtained receive beamforming vector, power allocation, phase shift matrix and ${\rm MSE}$ value at the $t$-th iteration, $T_{0,max}$ denotes the maximum iteration number and $\varepsilon_0$ denotes the convergence accuracy.

When CVX is adopted to solve subproblems, the interior-point algorithm is invoked by default. The worst-case complexity of {\bf{Algorithm 1}} is $\mathcal{O} ({T_{1,\max }}2K)$. Similarly, the worst-case complexity of solving $\mathcal{P}3.1$ and $\mathcal{P}4.3$ are $\mathcal{O} ({T_{2,\max }}{\left( {2K} \right)^3})$~\cite{Wanli2021Resource} and $\mathcal{O} ({M^{4.5}}\log \left( {{1 \mathord{\left/ {\vphantom {1 {{\varepsilon _3}}}} \right. \kern-\nulldelimiterspace} {{\varepsilon _3}}}} \right))$~\cite{Zhi2010Semidefinite}, respectively. Therefore, the overall worst-case complexity of {\bf{Algorithm 2}} can be obtained as $\mathcal{O} ({T_{0,\max }}{T_{1,\max }}2K + {T_{0,\max }}{T_{2,\max }}{\left( {2K} \right)^3} + {T_{0,\max }}{M^{4.5}}\log \left( {{1 \mathord{\left/ {\vphantom {1 {{\varepsilon _3}}}} \right. \kern-\nulldelimiterspace} {{\varepsilon _3}}}} \right))$.

\begin{algorithm}[t]
\caption{\emph{AO based Algorithm for Solving Problem ${\mathcal{P}1}$.}}
\label{AO based Algorithm for Solving Problem P1}
{\bf Input:}
feasible initial point $\left( {{{\bf{b}}^{\left( 0 \right)}},{{\bf{p}}^{\left( 0 \right)}},{{\bf{\Theta }}^{\left( 0 \right)}}} \right)$, $T_{0,max}$, $\varepsilon_0$.\\
{\bf Output:}
converged point $\left( {{{\bf{b}}^*},{{\bf{p}}^*},{{\bf{\Theta }}^*}} \right)$.

\begin{algorithmic}[1]
\STATE Initialize $T_{0,max}$, $\varepsilon_0$, $t=0$.
\WHILE {$t < {T_{0,max }}$ \&\& $\left| {MS{E^{\left( t \right)}} - MS{E^{\left( {t - 1} \right)}}} \right| > {\varepsilon _0}$ }
\STATE $t=t+1$.
\STATE Use ${{\bf{p}}^{\left( t-1 \right)}}$ and ${{\bf{\Theta }}^{\left( t-1 \right)}}$ to obtain ${{\bf{b}}^{\left( t \right)}}$ through {\bf{Algorithm 1}}.
\STATE Use ${{\bf{\Theta }}^{\left( t-1 \right)}}$ and ${{\bf{b}}^{\left( t \right)}}$ to obtain ${{\bf{p}}^{\left( t \right)}}$ through solving $\mathcal{P}3.1$ with CVX.
\STATE Use ${{\bf{b}}^{\left( t \right)}}$ and ${{\bf{p}}^{\left( t \right)}}$ to obtain ${{\bf{\Theta }}^{\left( t \right)}}$ through solving $\mathcal{P}4.3$ with CVX.
\ENDWHILE
\STATE $\left( {{{\bf{b}}^*},{{\bf{p}}^*},{{\bf{\Theta }}^*}} \right) = \left( {{{\bf{b}}^{\left( t \right)}},{{\bf{p}}^{\left( t \right)}},{{\bf{\Theta }}^{\left( t \right)}}} \right)$.
\STATE {\bf Return:} converged point $\left( {{{\bf{b}}^*},{{\bf{p}}^*},{{\bf{\Theta }}^*}} \right)$.
\end{algorithmic}
\end{algorithm}

\begin{figure*} [t]
	\begin{minipage}{0.33\textwidth}
		\centering
		\includegraphics[scale=0.44]{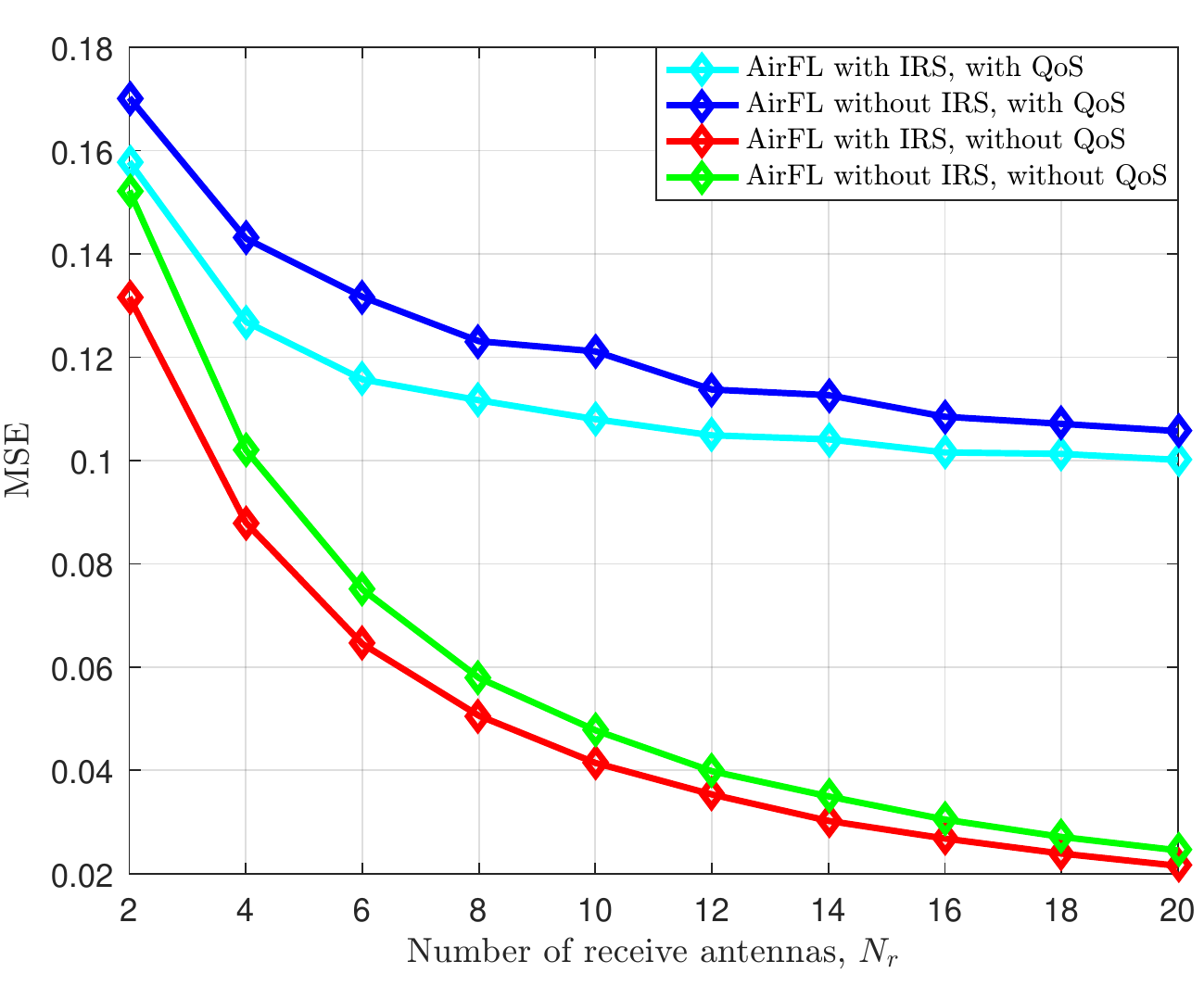}
		\centering
		\caption{MSE versus $N_r$.}
		\label{MSE versus Nr}
	\end{minipage}
	\begin{minipage}{0.33\textwidth}
		\centering
		\includegraphics[scale=0.44]{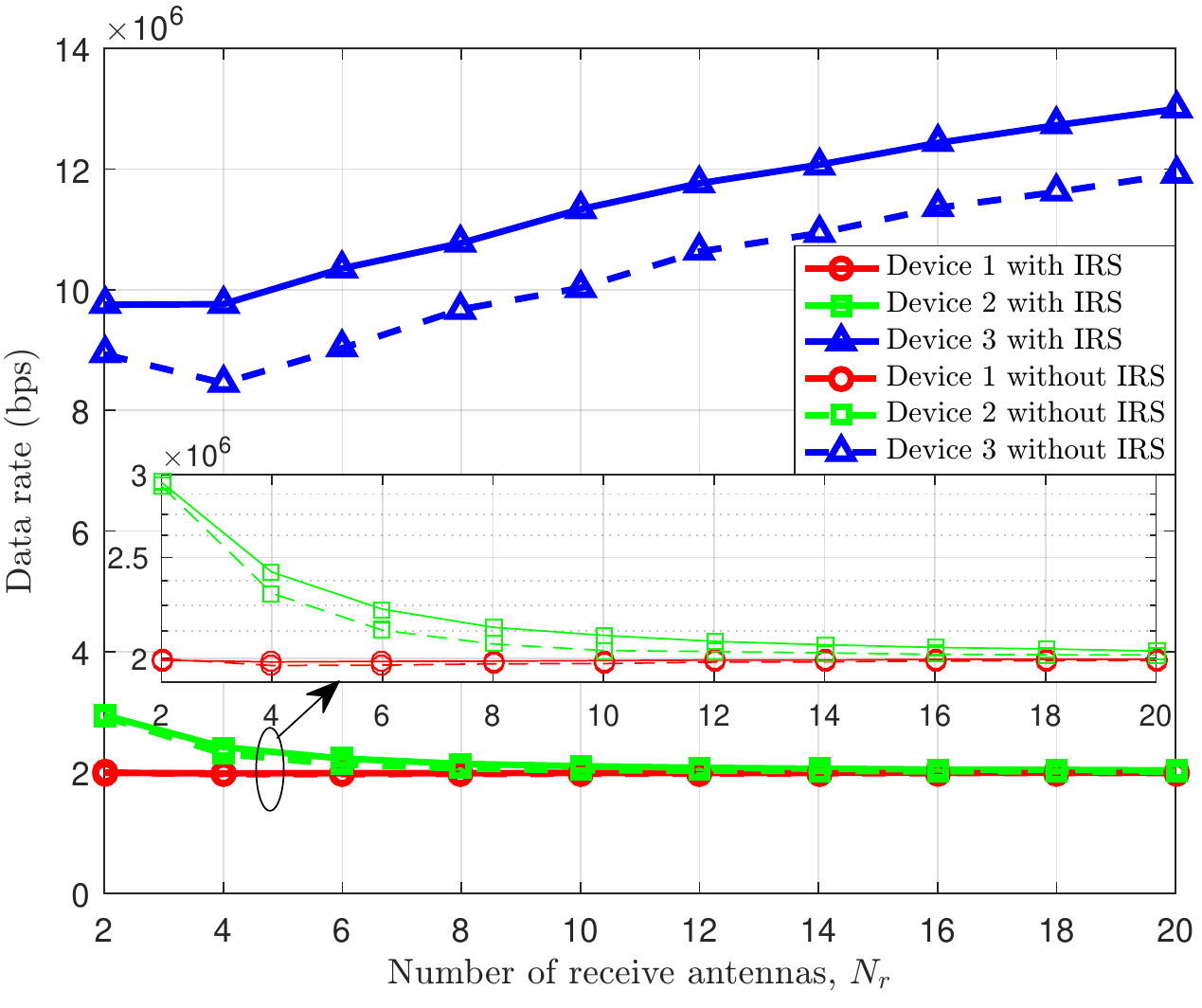}
		\centering
		\caption{Data rate versus $N_r$.}
		\label{MSE versus Nr}
	\end{minipage}
	\begin{minipage}{0.33\textwidth}
		\centering
		\includegraphics[scale=0.44]{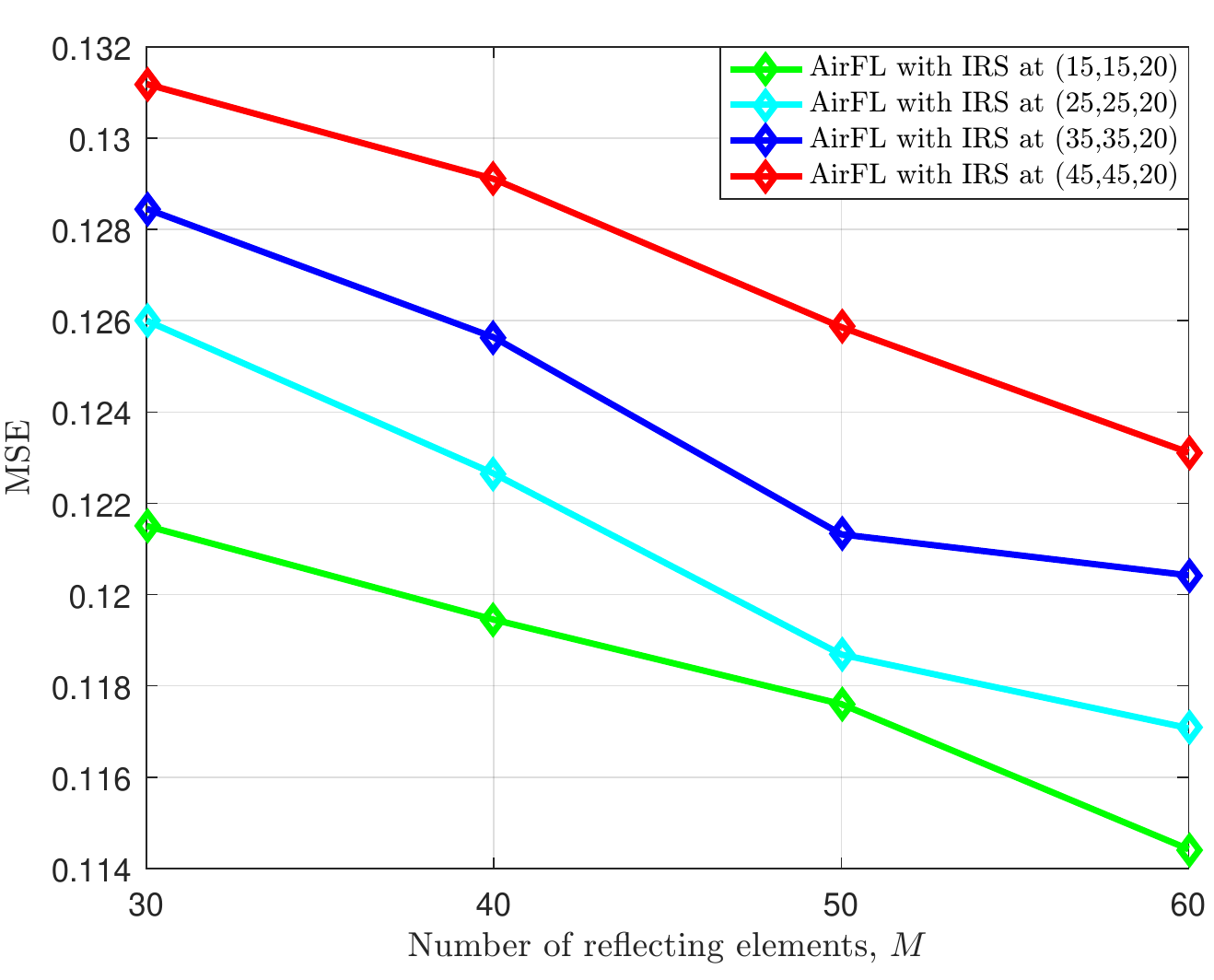}
		\centering
		\caption{MSE versus $M$.}
		\label{MSE versus M}
	\end{minipage}
\end{figure*}

\section{Simulation Results} \label{simulation results}
In this section, we validate the effectiveness of our proposed AO based algorithm. Unless specified otherwise, simulation parameters are set as follows. We assume an AirFL system consists of $3$ ground devices whose coordinates are randomly chosen from a $100$ m$\times$$100$ m square area and stay unchanged. The BS and the IRS with $30$ reflecting elements are located at $\left( 0,0,25 \right)$ and $\left( {25,25,20} \right)$ respectively. The available bandwidth $B=2$ MHz and QoS constraint $R_{\min } = 0.5$ Mbps. The maximum transit power is set as $P_{\max}=0$ dB and the minimum processed transmit power difference $p_{\rm gap}=10$ dBm. The AWGN power $\sigma ^2$ is $-80$ dBm. Other parameters are set as ${\varepsilon _0} = {\varepsilon _1} = {10^{ - 5}}$, ${T_{1,\max }} = {10^6}$, ${T_{0,\max }} = 40$ and ${T_{2,\max }}$ and ${\varepsilon _3}$ are set as the default values in CVX.

Fig 2 depicted the relationship between MSE and the number of receive antennas. As $N_r$ increases, one can obtain that the MSE monotonously decreases alongside, which demonstrated the fact that more receive antennas at BS effectively benefits the performance of AirFL. Meanwhile, AirFL systems with an IRS achieve lower MSE than those with no IRS. That is to say, IRS can promote the decrease of MSE by improving the quality of wireless channels. It is worth mentioning that our proposed algorithm suffers higher MSE compared with those schemes without QoS constraint. However, this reveals the truth that our proposed algorithm sacrifices the performance of MSE for satisfying the data rate constraint.

Fig 3 illustrated the achievable data rate under different number of receive antennas. It can be seen that all devices achieve higher data rate than $R_{\min}$. Concretely speaking, device $3$ achieves highest data rate in the sense that SIC has subtracted all interference while device $2$ possessing relatively lower rate due to interference from device $3$. However, the rate of device $1$ is the lowest among all devices owing to the strongest interference from other two devices. When $N_r$ is increasing, the rate of device $1$ and $3$ increases as well but device $2$ suffers lower rate due to increasing interference from device $3$. Furthermore, devices in AirFL systems with IRS outperform those without IRS in terms of data rate, which tells the truth that IRS contributes to higher data rate.

Fig 4 demonstrated the effect of the number of reflecting elements of the IRS on the MSE value. It is shown that more reflecting elements can lead to lower MSE for the reason that the wireless environment becomes more controllable and better channel conditions can be obtained with the increment of reflecting elements. Furthermore, it is found that for a given number of reflecting elements, MSE gradually increases as the location of IRS is setting farther from BS. This can be explained by the fact that the large-scale fading rises with the distance between the IRS and the BS.

\section{Conclusion} \label{conclusion}
In this paper, we investigated the QoS ensured MSE minimization problem in AirFL system assisted by IRS through jointly optimizing the receive beamforming vector, transmit power and phase shift matrix. To cope with the formulated intractable problem, we introduce alternating optimization to address the decomposed three subproblems in an iterative manner. Simulation results validate the availability of IRS and the effectiveness of our proposed algorithm in improving federated learning performance.

\bibliographystyle{IEEEtran}
\bibliography{IEEEabrv,reference}
	
\end{document}